%% file: main.tex
\documentclass[conference, 9pt]{IEEEtran}
\usepackage{epsfig,amssymb,amsmath}

\usepackage{multirow}
\usepackage[table,xcdraw]{xcolor}
\usepackage{microtype}
\usepackage[table]{xcolor}
\usepackage{todonotes}
\usepackage{url}
\usepackage{cleveref}
\usepackage{cite}
\usepackage{xspace}

\title{Easy, Interpretable,  Effective: 
openSMILE for voice deepfake detection
}

\author{\IEEEauthorblockN{Octavian Pascu}
\IEEEauthorblockA{
    \textit{\textsc{Politehnica} Bucharest}\\
    \texttt{octavian.pascu@upb.ro}
}
\and
\IEEEauthorblockN{Dan Oneață}
\IEEEauthorblockA{
    \textit{\textsc{Politehnica} Bucharest}\\
    \texttt{dan.oneata@gmail.com}
}
\and
\IEEEauthorblockN{Horia Cucu}
\IEEEauthorblockA{
    \textit{\textsc{Politehnica} Bucharest}\\
    \texttt{horia.cucu@upb.ro}
}
\and
\IEEEauthorblockN{Nicolas Müller}
\IEEEauthorblockA{
    \textit{\textsc{Fraunhofer AISEC}, Germany}\\
    \texttt{nicolas.mueller@aisec.fraunhofer.de}
}
}

\definecolor{mygreen}{HTML}{008000}
\definecolor{mypurple}{HTML}{9966CC}
\definecolor{myblue}{HTML}{5072A7}
\definecolor{mypapaya}{HTML}{EE892F}

\newcommand{\etal}{\textit{et al.}\xspace}
\newcommand{\train}{_{\mathrm{train}}}
\newcommand{\eval}{_{\mathrm{eval}}}

\begin{document}

\maketitle

\begin{abstract}




In this paper, we demonstrate that attacks in the latest ASVspoof5 dataset---a de facto standard in the field of voice authenticity and deepfake detection---can be identified with surprising accuracy using 
a small subset of very simplistic features.
These are derived from the openSMILE library,
and are scalar-valued, easy to compute, and human interpretable.
For example, attack $A10$'s unvoiced segments have a mean length of $0.09 \pm 0.02$, while bona fide instances have a mean length of $0.18 \pm 0.07$.
Using this feature alone, a threshold classifier achieves an Equal Error Rate (EER) of $10.3\%$ for attack $A10$.
Similarly, across all attacks, we achieve up to $0.8\%$ EER, with an overall EER of $15.7 \pm 6.0$\%.


We explore the generalization capabilities of these features and find that some of them transfer effectively between attacks, primarily when the attacks originate from similar Text-to-Speech (TTS) architectures.
This finding may indicate that voice anti-spoofing is, in part, a problem of identifying and remembering signatures or fingerprints of individual TTS systems.
This allows to better understand anti-spoofing models and their challenges in real-world application. 



\end{abstract}






\section{Introduction}

Text-to-speech (TTS) technology has advanced significantly in recent years, offering various beneficial applications, such as the restoration of voice capabilities for speech-impaired individuals \cite{biadsy2019parrotron}. Despite these positive developments, TTS also presents potential risks such as the compromise of voice biometric systems and the creation of deepfakes, which enable fraud, slander and misinformation. 
A key initiative in combating these threats is the ASVspoof Challenge \cite{wu2017asvspoof}. 
Since its inception in 2015, this biennial event has released datasets crucial for the design of anti-spoofing systems. 
Notably, since the forth iteration of ASVspoof in 2021~\cite{yamagishi2021asvspoof, asvspoof2021taslp}, the challenge has included a separate track specifically for detecting audio deepfakes.
The latest iteration of the ASVspoof challenge dataset is called `ASVspoof 5', and is set to succeed the ASVspoof 2021 dataset, the de-facto standard in evaluating of voice authenticity systems.

While related work reports high performance on the ASVspoof datasets~\cite{tak2021end,jung2022aasist,kawa2023improved}
, achieving true generalization that applies effectively in real-world scenarios remains a challenge. 
Existing studies suggest that anti-spoofing efforts often rely on shortcut artifacts, such as the length of silences \cite{muller2021speech}
or bitrate information \cite{borzi2022synthetic}, and struggle with cross-dataset generalization \cite{muller2022generalize}.
Large self-supervised representations improve the generalization to some degree \cite{pascu24interspeech}, but this approach comes at the cost of explainability, which remains an important desideratum given the decision-critical nature of deepfake detection.

\begin{figure}[t]
    \centering
    \includegraphics[width=0.995\linewidth]{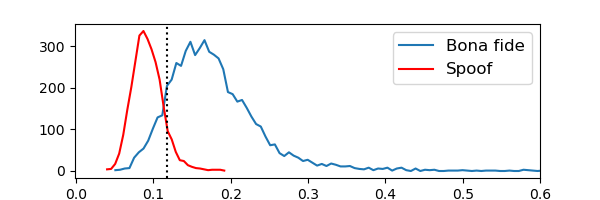}
    \caption{Distribution of the openSMILE `eGeMAPSv2' feature F85, `MeanUnvoicedSegmentLength', computed for attack $A10$ and bona fide data from ASVspoof5. 
    A simple threshold classifier obtains an EER of $10.3\%$ by predicting `spoof' if $F85 < 0.12$, else `bona fide' (dotted line).}
    \label{fig:dist}
\end{figure}

In this work, we propose to use openSMILE \cite{eyben2010opensmile}, a software tool for the automatic extraction of audio features. 
This tool allows the computation of interpretable features, based on fundamental properties such as the length of voiced and unvoiced segments, spectral flux (a measure of how quickly the power spectrum of a signal is changing), or energy within specific frequency ranges. 
Our findings demonstrate that for the ASVspoof5 dataset, even single, scalar-valued openSMILE features allow for surprisingly accurate classification of attacks (i.e. voice deepfakes from a specific TTS system). 
For example, the `MeanUnvoiceSegmentLength' feature allows for reliable identification of attack $A10$, c.f. \Cref{fig:dist}.
We find that for each attack in ASVspoof5, such features can be found.
Interestingly, the features extracted exhibit varying degrees of specificity: 
some are highly specific and fail to generalize across different types of attacks, while others show broader applicability. 
This variation suggests that the underlying text-to-speech (TTS) models each possess unique characteristics that, if recognized during training, facilitate identification.
In summary:
\begin{itemize}
    \item We employ openSMILE to identify single, scalar-valued and human-interpretable features that effectively pinpoint attacks within the ASVspoof5 dataset.
    \item We show that cross-domain generalization works best when attacks come from similar TTS architectures.
    \item This observation suggests that TTS models exhibit unique characteristics, akin to fingerprints, which are readily identifiable if seen during training, but can pose significant challenges in cross-model generalization.

\end{itemize}

\begin{figure*}
    \centering
    \includegraphics[width=0.75\linewidth]{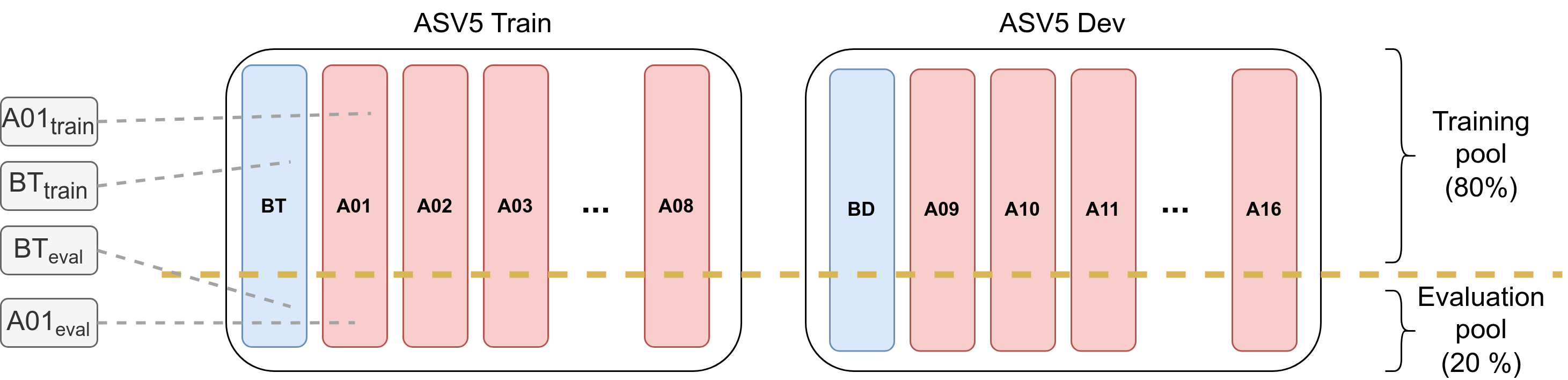}
    \caption{Visualization of the ASVspoof5 dataset's `train' and `dev' partition.
    BT and BD correspond to the respective bona fide data, while $A01$ through $A16$ correspond to the individual attacks. Grey boxes on the left indicate the naming convention we use in subsequent experiments.}
    \label{fig:data_pools}
\end{figure*}

\section{Methodology}
\input{tex/methodology}

\section{Experiments}
\input{tex/experiments/expA}

\input{tex/results/expA_2feats}

\input{tex/asv_systems}
\input{tex/experiments/expB}
\input{tex/results/expB_2feats}

\input{tex/experiments/expC}

\input{tex/results/expC_single}
\input{tex/results/expC_multi}

\section{Related Work}
\input{tex/related_work}

\section{Discussion and Conclusion}
In this work, we analyze the efficacy of openSMILE as a front-end for voice anti-spoofing, specifically on the ASVspoof5 dataset.
We find that for all attacks, there exist scalar-valued, human interpretable features which allow for good in-domain detection performance.
However, the application of these characteristics in out-of-domain settings remains limited, particularly across different TTS architectures.

We argue that TTS models display unique characteristics, akin to a fingerprint, which can be readily identified once encountered but can have limited use for generalization across different models.
Judging by the good transferability between attacks originating from similar TTS architectures, c.f. \Cref{tab:attack_desc} and \Cref{tab:expB}, these characteristics appear to include information about the underlying TTS architecture.
Conversely, generalization across architecture boundaries is challenging.
Success in the field of source tracing \cite{klein2024source, muller22attacker}, which involves assigning instances to their generating models, further demonstrates that individual models exhibit individual characteristics.
Thus, voice anti-spoofing may, in part, involve the challenge of identifying the unique signatures or fingerprints of individual TTS systems. 
While these distinctive characteristics can enable  good identification of an attack, they require that the TTS model has been encountered during training.
This may in part explain challenges in real-world application~\cite{muller2022generalize}.

\bibliographystyle{IEEEtran}
\bibliography{main}

\end{document}

%% file: tex/methodology.tex
\subsection{Data Partitioning}
We utilize the two currently available, labeled partitions of the ASVspoof5~\cite{Wang2024_ASVspoof5} dataset: `train' and `dev'.
Each partition a set of bona fide instances, denoted as $BT$ and $BD$ (bona fide `train' and bona fide `dev', respectively). 
Additionally, each partition includes spoofed audio files, categorized by the TTS system used to generate them, referred to as `attacks'. 
The `train' partition comprises attacks $A01$ through $A08$, while the `dev' partition includes attacks $A09$ through $A16$; 
the types of attacks are listed in \Cref{tab:attack_desc}.
For our experiments, we divided $BT$ and $BD$, along with each attack, into an 80\% training pool and a 20\% evaluation pool, c.f. 
\Cref{fig:data_pools}.

\subsection{Identifying Predictive openSMILE Features}\label{meth:how_to_find_feats}

Our objective is to ascertain whether attacks in ASVspoof5 can be detected using a single, scalar-valued feature from openSMILE. 
We aim to evaluate the performance of this method in both in-domain (ID) and out-of-domain (OOD) scenarios. 
ID scenarios involve training and evaluation data from the same attack, whereas OOD scenarios involve training and evaluation data from different attacks. 
For instance, consider a classifier trained on bona fide instances from the training pool ($BD\train$ and $BT\train$), as well as spoofed instances from attack $A01\train$. 
Evaluating performance on $BD\eval$, $BT\eval$, and $A01\eval$ constitutes an ID challenge, while evaluating performance on $BD\eval$, $BT\eval$, and $A02\eval$ constitutes an OOD challenge. 
Prior research~\cite{muller2022generalize,cuccovillo2022wifs,yi2023arxiv}
highlights that OOD generalization is challenging but essential for real-world deepfake detection applications.

We aim to identify the most predictive features for each attack $Aj$.
To this end, we utilize openSMILE's `eGmapsV2' feature set \cite{eyben2016egemaps}, consisting of $88$ scalar-valued features derived from an audio's loudness, spectral flux, voiced and unvoiced segment lengths, etc.
This set is an extended version of the Geneva Minimalistic Acoustic Parameter (eGeMAPS) feature set, originally employed for emotion recognition and affective computing.
While we have considered other feature sets, such as `emobase', we found them to be less interpretable.

To illustrate our approach, consider \Cref{fig:dist}. 
The plot shows the distribution of the feature `MeanUnvoicedSegmentLength' (shorthand $F85$, being the 86th feature from `eGeMAPSv2') for bona fide instances and attack $A10$.
The red line represents spoofed data from $A10$,
while the blue line represents bona fide data from the `train' and `dev' splits. 
It is evident that this feature exhibits distinct behavior between spoofed and real data. 
For spoofed data, this feature assumes values of $0.09 \pm 0.02$ (mean and standard deviation, respectively), while for bona fide data, it assumes values of $0.18 \pm 0.07$.
Using this feature as a predictor, a threshold classifier achieves an EER of 10.3\%, much better than random guessing (50\% EER). 
Therefore, `MeanUnvoicedSegmentLength' is a good predictor for attack $A10$. 
In subsequent experiments, we find the most predictive features using this approach: 
For each attack $Aj$ and each `eGeMAPSv2' feature, compute the EER on $BT_{\text{train}}$, $BD_{\text{train}}$, and $Aj\train$. Then, we present the two features that individually yield the best EER.

%% file: tex/experiments/expA.tex
\subsection{Experiment A: openSMILE, In-domain}\label{sec:expA}

In this initial experiment, we evaluate the effectiveness of `eGeMAPSv2' 
in an in-domain (ID) setting, where the training and testing data originate from the same distribution.

First, we select the front-end: as outlined in \Cref{meth:how_to_find_feats}, we identify the two most predictive features per attack.
Second, we train the back-end on each feature individually: A linear classification model \( f(x) = wx + b \in \mathbb{R} \) where \( x, w, b \in \mathbb{R} \). 
This model is trained on the `train' portion of the attack data along with the bona fide training data (i.e. $BD\train$, $BT\train$ and $Aj\train$, c.f. \Cref{fig:data_pools}).
We then compute the EER on the corresponding evaluation data, i.e. $BD\eval$, $BT\eval$ and $Aj\eval$.

%% file: tex/results/expA_2feats.tex

\Cref{tab:expA} presents the results for in-domain evaluation.
For most attacks, a single openSMILE feature is sufficient to achieve good performance, with results as low as $0.8\%$ EER for attack $A14$. The worst performance is observed for attack $A16$, where the best performing feature $F66$ yields an EER of $23.8\%$. 
On average, this simplistic approach achieves $15.7 \pm 6.0$\%. EER---much better than random guessing.

These results indicate that each attack has a distinctive fingerprint, and is characterized by unique, but simplistic patterns. 
While some attacks are more easily identified than others, this principle holds throughout ASVspoof5. 
Interestingly, certain features, such as $F19$, are effective across multiple attacks (7 out of 16 attacks), whereas others, like $F51$, are highly effective for specific attacks (e.g., $A14$ with $0.8\%$ EER) but not for others.
Some of the most occurring features in \Cref{tab:expA} are:
\begin{itemize}
\itemsep0em
    \item \textbf{F19 (7 occurrences):} loudness\_sma3\_stddevFallingSlope -- standard deviation of the falling slope of the simple moving average (with a window size of 3) of the loudness.
    \item \textbf{F66 (6 occurrences):} spectralFluxV\_sma3nz\_amean -- arithmetic mean of the simple moving average (with a window size of 3) of the spectral flux for non-zero values.
    \item \textbf{F85 (2 occurrences):} MeanUnvoicedSegmentLength -- average length of all unvoiced segments within the audio file.
    \item \textbf{F86 (2 occurrences):} StddevUnvoicedSegmentLength -- standard deviation of the length of all unvoiced segments.
\end{itemize}

\begin{table}[]
\centering
\caption{In-domain test performance per attack, for the top-2 features per attack $A01$ through $A16$.}
\resizebox{\columnwidth}{!}{%
\begin{tabular}{|c|c|c|c|c|c|c|}
\cline{1-3} \cline{5-7}
Attack & Feature & \multicolumn{1}{l|}{Test EER} &  & Attack & Feature & \multicolumn{1}{l|}{Test EER} \\ \cline{1-3} \cline{5-7} 
\multirow{2}{*}{A01} & F66 & 12.8 &  & \multirow{2}{*}{A09} & F86 &  8.0 \\ \cline{2-3} \cline{6-7} 
                     & F19 & 13.8 &  &                      & F85 & 11.5 \\ \cline{1-3} \cline{5-7} 
\multirow{2}{*}{A02} & F66 & 12.7 &  & \multirow{2}{*}{A10} & F86 &  7.3 \\ \cline{2-3} \cline{6-7} 
                     & F19 & 13.0 &  &                      & F85 & 10.7 \\ \cline{1-3} \cline{5-7} 
\multirow{2}{*}{A03} & F19 & 15.3 &  & \multirow{2}{*}{A11} & F58 & 15.3 \\ \cline{2-3} \cline{6-7} 
                     & F66 & 15.7 &  &                      & F60 & 15.3 \\ \cline{1-3} \cline{5-7} 
\multirow{2}{*}{A04} & F19 & 16.8 &  & \multirow{2}{*}{A12} & F82 & 18.2 \\ \cline{2-3} \cline{6-7} 
                     & F21 & 17.3 &  &                      & F19 & 21.3 \\ \cline{1-3} \cline{5-7} 
\multirow{2}{*}{A05} & F21 & 17.5 &  & \multirow{2}{*}{A13} & F67 & 22.1 \\ \cline{2-3} \cline{6-7} 
                     & F19 & 19.4 &  &                      & F39 & 22.8 \\ \cline{1-3} \cline{5-7} 
\multirow{2}{*}{A06} & F21 & 16.8 &  & \multirow{2}{*}{A14} & F45 &  0.8 \\ \cline{2-3} \cline{6-7} 
                     & F19 & 17.1 &  &                      & F51 &  0.8 \\ \cline{1-3} \cline{5-7} 
\multirow{2}{*}{A07} & F66 & 13.7 &  & \multirow{2}{*}{A15} & F78 & 20.7 \\ \cline{2-3} \cline{6-7} 
                     & F1  & 14.2 &  &                      & F77 & 25.7 \\ \cline{1-3} \cline{5-7} 
\multirow{2}{*}{A08} & F20 & 17.7 &  & \multirow{2}{*}{A16} & F66 & 23.8 \\ \cline{2-3} \cline{6-7} 
                     & F66 & 18.2 &  &                      & F18 & 26.8 \\ \cline{1-3} \cline{5-7} 
\end{tabular}%
}
\label{tab:expA}
\end{table}



%% file: tex/asv_systems.tex
\begin{table}[t]
    \centering
    \caption{TTS system per spoofing attack, as presented in \cite{Wang2024_ASVspoof5}.}
    \begin{tabular}{c|c||c|c}
         Attack & System & Attack & System \\ \hline
         A01 & GlowTTS \cite{kim2020glow} & A09 & ToucanTTS \cite{lux2022low} \\
         A02 & Variant of A01& A10& A09+HifiGANv2 \cite{kong2020hifi} \\
         A03 & Variant of A01 & A11& Tacotron2 \cite{shen2018natural}\\
         A04 & GradTTS \cite{popov2021grad} & A12& In-house unit-select\\
         A05 & Variant of A04& A13& StarGANv2-VC \cite{li2021starganv2}  \\
         A06 & Variant of A04 & A14& YourTTS \cite{casanova2022yourtts}\\
         A07 & FastPitch \cite{lancucki2021fastpitch} & A15& VAE-GAN \cite{albadawy2020voice}\\
         A08 & VITS \cite{kim2021conditional} & A16& In-house ASR-based
    \end{tabular}
    \label{tab:attack_desc}
\end{table}

%% file: tex/experiments/expB.tex
\subsection{Experiment B: openSMILE, Out-of-domain}
\label{subsec:exp-b}

In our second experiment, we extend our approach to an out-of-domain (OOD) setting, where the training and testing data come from different distributions. 
This aims to evaluate the effectiveness of openSMILE features for cross-domain detection in the ASVspoof5 dataset.
We follow the same training protocol as in \Cref{sec:expA}, but evaluate on OOD data. 
For instance, given a model trained on $A01$, we evaluate on $A02$ through $A16$ and present the EERs individually. 
Note that both the feature itself, as well as the parameters $w, b \in \mathbb{R}$ are derived from the initial in-domain training.
This approach results in a matrix of OOD performances, where the entries on the diagonal correspond to the in-domain (ID) values, as shown in \Cref{tab:expA}.

%% file: tex/results/expB_2feats.tex

The results are shown in \Cref{tab:expB}, where the rows indicate the training attack, and the columns indicate the evaluation attack.
Note that \Cref{tab:expB} can be partitioned into four sub-matrices based on the `train' and `dev' partitions of the ASVspoof 5 dataset.
Systems trained on attacks $A01$ through $A08$
generalize relatively well to other `train' attacks, as indicated by the green highlights and correspondingly lower EERs ($20.0 \pm 7.0$\%) in the top-left quadrant. 

We observe that certain groups of attacks exhibit strong generalization between themselves, such as $A09$ and $A10$ ($9.3 \pm 1.7\% EER)$. 
We hypothesize that this generalization is due to similarities in the underlying TTS algorithms (see \Cref{tab:attack_desc} for details). 
For example, both $A09$ and $A10$ are derived from ToucanTTS \cite{lux2022low}, which explains their effective generalization. 
Similarly, $A01$, $A02$, and $A03$ are based on the GlowTTS architecture \cite{kim2020glow}, which likely contributes to the strong generalization observed among them.
Generalization across different TTS architectures often proves challenging; for example, systems trained on $A16$ do not generalize to any of the attacks from $A01$ through $A08$.

In summary, generalization from `dev' to `train' and vice-versa is challenging, as indicated by the predominantly red highlights and high EERs ($52.9 \pm 17.9$\%) in the bottom-left and top-right quadrants.



\begin{table*}[]
\centering
\caption{Out-of-domain evaluation results presented in terms of EER.
Rows indicate
training attack and openSMILE feature used, while columns indicate evaluation attack.
Note that the values on the diagonal originate from identical train and test attacks, and thus represent in-domain evaluation. 
}
\resizebox{\linewidth}{!}{%
\begin{tabular}{|l|l|rrrrrrrr|rrrrrrrr}
\hline
  %
  & & \multicolumn{8}{c|}{ASVspoof 5 `train'} & \multicolumn{8}{c|}{ASVspoof 5 `dev'} \\
\hline
 &
   &
  \multicolumn{1}{l|}{} &
  \multicolumn{1}{l|}{} &
  \multicolumn{1}{l|}{} &
  \multicolumn{1}{l|}{} &
  \multicolumn{1}{l|}{} &
  \multicolumn{1}{l|}{} &
  \multicolumn{1}{l|}{} &
  \multicolumn{1}{l|}{} &
  \multicolumn{1}{l|}{} &
  \multicolumn{1}{l|}{} &
  \multicolumn{1}{l|}{} &
  \multicolumn{1}{l|}{} &
  \multicolumn{1}{l|}{} &
  \multicolumn{1}{l|}{} &
  \multicolumn{1}{l|}{} &
  \multicolumn{1}{l|}{} \\
 &
   &
  \multicolumn{1}{l|}{} &
  \multicolumn{1}{l|}{} &
  \multicolumn{1}{l|}{} &
  \multicolumn{1}{l|}{} &
  \multicolumn{1}{l|}{} &
  \multicolumn{1}{l|}{} &
  \multicolumn{1}{l|}{} &
  \multicolumn{1}{l|}{} &
  \multicolumn{1}{l|}{} &
  \multicolumn{1}{l|}{} &
  \multicolumn{1}{l|}{} &
  \multicolumn{1}{l|}{} &
  \multicolumn{1}{l|}{} &
  \multicolumn{1}{l|}{} &
  \multicolumn{1}{l|}{} &
  \multicolumn{1}{l|}{} \\
\multirow{-3}{*}{Attack $\downarrow$} &
  \multirow{-3}{*}{Feature $\downarrow$} &
  \multicolumn{1}{l|}{\multirow{-3}{*}{A01}} &
  \multicolumn{1}{l|}{\multirow{-3}{*}{A02}} &
  \multicolumn{1}{l|}{\multirow{-3}{*}{A03}} &
  \multicolumn{1}{l|}{\multirow{-3}{*}{A04}} &
  \multicolumn{1}{l|}{\multirow{-3}{*}{A05}} &
  \multicolumn{1}{l|}{\multirow{-3}{*}{A06}} &
  \multicolumn{1}{l|}{\multirow{-3}{*}{A07}} &
  \multicolumn{1}{l|}{\multirow{-3}{*}{A08}} &
  \multicolumn{1}{l|}{\multirow{-3}{*}{A09}} &
  \multicolumn{1}{l|}{\multirow{-3}{*}{A10}} &
  \multicolumn{1}{l|}{\multirow{-3}{*}{A11}} &
  \multicolumn{1}{l|}{\multirow{-3}{*}{A12}} &
  \multicolumn{1}{l|}{\multirow{-3}{*}{A13}} &
  \multicolumn{1}{l|}{\multirow{-3}{*}{A14}} &
  \multicolumn{1}{l|}{\multirow{-3}{*}{A15}} &
  \multicolumn{1}{l|}{\multirow{-3}{*}{A16}} \\ \hline
 &
  F66 &
  \cellcolor[HTML]{86C280}12.8 &
  \cellcolor[HTML]{86C280}12.7 &
  \cellcolor[HTML]{92C47E}15.7 &
  \cellcolor[HTML]{A9C879}21.7 &
  \cellcolor[HTML]{D0CE70}31.5 &
  \cellcolor[HTML]{B9CA75}25.7 &
  \cellcolor[HTML]{8AC380}13.7 &
  \cellcolor[HTML]{9BC67C}18.2 &
  \cellcolor[HTML]{E67C73}68.7 &
  \cellcolor[HTML]{E67C73}69.5 &
  \cellcolor[HTML]{E67C73}68.1 &
  \cellcolor[HTML]{E67C73}70.3 &
  \cellcolor[HTML]{E67C73}60.0 &
  \cellcolor[HTML]{E67C73}55.2 &
  \cellcolor[HTML]{E67C73}68.7 &
  \cellcolor[HTML]{E67C73}76.2 \\ \cline{2-2}
\multirow{-2}{*}{A01} &
  F19 &
  \cellcolor[HTML]{8AC37F}13.8 &
  \cellcolor[HTML]{87C280}13.0 &
  \cellcolor[HTML]{90C47E}15.3 &
  \cellcolor[HTML]{96C57D}16.8 &
  \cellcolor[HTML]{A0C67B}19.4 &
  \cellcolor[HTML]{97C57D}17.1 &
  \cellcolor[HTML]{9AC57C}17.8 &
  \cellcolor[HTML]{AFC978}23.0 &
  \cellcolor[HTML]{E67C73}61.9 &
  \cellcolor[HTML]{E67C73}62.8 &
  \cellcolor[HTML]{E67C73}55.7 &
  \cellcolor[HTML]{E67C73}78.7 &
  \cellcolor[HTML]{E67C73}54.7 &
  \cellcolor[HTML]{E67C73}54.5 &
  \cellcolor[HTML]{E67C73}65.3 &
  \cellcolor[HTML]{E67C73}66.7 \\ \cline{1-2}
  &
  F66 &
  \cellcolor[HTML]{86C280}12.8 &
  \cellcolor[HTML]{86C280}12.7 &
  \cellcolor[HTML]{92C47E}15.7 &
  \cellcolor[HTML]{A9C879}21.7 &
  \cellcolor[HTML]{D0CE70}31.5 &
  \cellcolor[HTML]{B9CA75}25.7 &
  \cellcolor[HTML]{8AC380}13.7 &
  \cellcolor[HTML]{9BC67C}18.2 &
  \cellcolor[HTML]{E67C73}68.7 &
  \cellcolor[HTML]{E67C73}69.5 &
  \cellcolor[HTML]{E67C73}68.1 &
  \cellcolor[HTML]{E67C73}70.3 &
  \cellcolor[HTML]{E67C73}60.0 &
  \cellcolor[HTML]{E67C73}55.2 &
  \cellcolor[HTML]{E67C73}68.7 &
  \cellcolor[HTML]{E67C73}76.2 \\ \cline{2-2}
\multirow{-2}{*}{A02} &
  F19 &
  \cellcolor[HTML]{8AC37F}13.8 &
  \cellcolor[HTML]{87C280}13.0 &
  \cellcolor[HTML]{90C47E}15.3 &
  \cellcolor[HTML]{96C57D}16.8 &
  \cellcolor[HTML]{A0C67B}19.4 &
  \cellcolor[HTML]{97C57D}17.1 &
  \cellcolor[HTML]{9AC57C}17.8 &
  \cellcolor[HTML]{AFC978}23.0 &
  \cellcolor[HTML]{E67C73}61.9 &
  \cellcolor[HTML]{E67C73}62.8 &
  \cellcolor[HTML]{E67C73}55.7 &
  \cellcolor[HTML]{E67C73}78.7 &
  \cellcolor[HTML]{E67C73}54.7 &
  \cellcolor[HTML]{E67C73}54.5 &
  \cellcolor[HTML]{E67C73}65.3 &
  \cellcolor[HTML]{E67C73}66.7 \\ \cline{1-2}
  &
  F19 &
  \cellcolor[HTML]{8AC37F}13.8 &
  \cellcolor[HTML]{87C280}13.0 &
  \cellcolor[HTML]{90C47E}15.3 &
  \cellcolor[HTML]{96C57D}16.8 &
  \cellcolor[HTML]{A0C67B}19.4 &
  \cellcolor[HTML]{97C57D}17.1 &
  \cellcolor[HTML]{9AC57C}17.8 &
  \cellcolor[HTML]{AFC978}23.0 &
  \cellcolor[HTML]{E67C73}61.9 &
  \cellcolor[HTML]{E67C73}62.8 &
  \cellcolor[HTML]{E67C73}55.7 &
  \cellcolor[HTML]{E67C73}78.7 &
  \cellcolor[HTML]{E67C73}54.7 &
  \cellcolor[HTML]{E67C73}54.5 &
  \cellcolor[HTML]{E67C73}65.3 &
  \cellcolor[HTML]{E67C73}66.7 \\ \cline{2-2}
 &
  F66 &
  \cellcolor[HTML]{86C280}12.8 &
  \cellcolor[HTML]{86C280}12.7 &
  \cellcolor[HTML]{92C47E}15.7 &
  \cellcolor[HTML]{A9C879}21.7 &
  \cellcolor[HTML]{D0CE70}31.5 &
  \cellcolor[HTML]{B9CA75}25.7 &
  \cellcolor[HTML]{8AC380}13.7 &
  \cellcolor[HTML]{9BC67C}18.2 &
  \cellcolor[HTML]{E67C73}68.7 &
  \cellcolor[HTML]{E67C73}69.5 &
  \cellcolor[HTML]{E67C73}68.1 &
  \cellcolor[HTML]{E67C73}70.3 &
  \cellcolor[HTML]{E67C73}60.0 &
  \cellcolor[HTML]{E67C73}55.2 &
  \cellcolor[HTML]{E67C73}68.7 &
  \cellcolor[HTML]{E67C73}76.2 \\ \cline{1-2}
\multirow{-3}{*}{A03} &
  F19 &
  \cellcolor[HTML]{8AC37F}13.8 &
  \cellcolor[HTML]{87C280}13.0 &
  \cellcolor[HTML]{90C47E}15.3 &
  \cellcolor[HTML]{96C57D}16.8 &
  \cellcolor[HTML]{A0C67B}19.4 &
  \cellcolor[HTML]{97C57D}17.1 &
  \cellcolor[HTML]{9AC57C}17.8 &
  \cellcolor[HTML]{AFC978}23.0 &
  \cellcolor[HTML]{E67C73}61.9 &
  \cellcolor[HTML]{E67C73}62.8 &
  \cellcolor[HTML]{E67C73}55.7 &
  \cellcolor[HTML]{E67C73}78.7 &
  \cellcolor[HTML]{E67C73}54.7 &
  \cellcolor[HTML]{E67C73}54.5 &
  \cellcolor[HTML]{E67C73}65.3 &
  \cellcolor[HTML]{E67C73}66.7 \\ \cline{2-2}
 &
  F21 &
  \cellcolor[HTML]{AAC879}21.9 &
  \cellcolor[HTML]{A9C879}21.6 &
  \cellcolor[HTML]{B6CA76}24.8 &
  \cellcolor[HTML]{98C57C}17.3 &
  \cellcolor[HTML]{99C57C}17.5 &
  \cellcolor[HTML]{96C57D}16.8 &
  \cellcolor[HTML]{9DC67B}18.7 &
  \cellcolor[HTML]{F6B66A}45.7 &
  \cellcolor[HTML]{9BC67C}18.1 &
  \cellcolor[HTML]{99C57C}17.5 &
  \cellcolor[HTML]{F2D369}40.0 &
  \cellcolor[HTML]{FED066}43.7 &
  \cellcolor[HTML]{D9D06E}33.8 &
  \cellcolor[HTML]{C5CC73}28.6 &
  \cellcolor[HTML]{E4D16C}36.5 &
  \cellcolor[HTML]{F8D468}41.5 \\ \cline{1-2}
\multirow{-3}{*}{A04} &
  F21 &
  \cellcolor[HTML]{AAC879}21.9 &
  \cellcolor[HTML]{A9C879}21.6 &
  \cellcolor[HTML]{B6CA76}24.8 &
  \cellcolor[HTML]{98C57C}17.3 &
  \cellcolor[HTML]{99C57C}17.5 &
  \cellcolor[HTML]{96C57D}16.8 &
  \cellcolor[HTML]{9DC67B}18.7 &
  \cellcolor[HTML]{F6B66A}45.7 &
  \cellcolor[HTML]{9BC67C}18.1 &
  \cellcolor[HTML]{99C57C}17.5 &
  \cellcolor[HTML]{F2D369}40.0 &
  \cellcolor[HTML]{FED066}43.7 &
  \cellcolor[HTML]{D9D06E}33.8 &
  \cellcolor[HTML]{C5CC73}28.6 &
  \cellcolor[HTML]{E4D16C}36.5 &
  \cellcolor[HTML]{F8D468}41.5 \\ \cline{2-2}
 &
  F19 &
  \cellcolor[HTML]{8AC37F}13.8 &
  \cellcolor[HTML]{87C280}13.0 &
  \cellcolor[HTML]{90C47E}15.3 &
  \cellcolor[HTML]{96C57D}16.8 &
  \cellcolor[HTML]{A0C67B}19.4 &
  \cellcolor[HTML]{97C57D}17.1 &
  \cellcolor[HTML]{9AC57C}17.8 &
  \cellcolor[HTML]{AFC978}23.0 &
  \cellcolor[HTML]{E67C73}61.9 &
  \cellcolor[HTML]{E67C73}62.8 &
  \cellcolor[HTML]{E67C73}55.7 &
  \cellcolor[HTML]{E67C73}78.7 &
  \cellcolor[HTML]{E67C73}54.7 &
  \cellcolor[HTML]{E67C73}54.5 &
  \cellcolor[HTML]{E67C73}65.3 &
  \cellcolor[HTML]{E67C73}66.7 \\ \cline{1-2}
\multirow{-3}{*}{A05} &
  F21 &
  \cellcolor[HTML]{AAC879}21.9 &
  \cellcolor[HTML]{A9C879}21.6 &
  \cellcolor[HTML]{B6CA76}24.8 &
  \cellcolor[HTML]{98C57C}17.3 &
  \cellcolor[HTML]{99C57C}17.5 &
  \cellcolor[HTML]{96C57D}16.8 &
  \cellcolor[HTML]{9DC67B}18.7 &
  \cellcolor[HTML]{F6B66A}45.7 &
  \cellcolor[HTML]{9BC67C}18.1 &
  \cellcolor[HTML]{99C57C}17.5 &
  \cellcolor[HTML]{F2D369}40.0 &
  \cellcolor[HTML]{FED066}43.7 &
  \cellcolor[HTML]{D9D06E}33.8 &
  \cellcolor[HTML]{C5CC73}28.6 &
  \cellcolor[HTML]{E4D16C}36.5 &
  \cellcolor[HTML]{F8D468}41.5 \\ \cline{2-2}
 &
  F19 &
  \cellcolor[HTML]{8AC37F}13.8 &
  \cellcolor[HTML]{87C280}13.0 &
  \cellcolor[HTML]{90C47E}15.3 &
  \cellcolor[HTML]{96C57D}16.8 &
  \cellcolor[HTML]{A0C67B}19.4 &
  \cellcolor[HTML]{97C57D}17.1 &
  \cellcolor[HTML]{9AC57C}17.8 &
  \cellcolor[HTML]{AFC978}23.0 &
  \cellcolor[HTML]{E67C73}61.9 &
  \cellcolor[HTML]{E67C73}62.8 &
  \cellcolor[HTML]{E67C73}55.7 &
  \cellcolor[HTML]{E67C73}78.7 &
  \cellcolor[HTML]{E67C73}54.7 &
  \cellcolor[HTML]{E67C73}54.5 &
  \cellcolor[HTML]{E67C73}65.3 &
  \cellcolor[HTML]{E67C73}66.7 \\ \cline{1-2}
\multirow{-3}{*}{A06} &
  F66 &
  \cellcolor[HTML]{86C280}12.8 &
  \cellcolor[HTML]{86C280}12.7 &
  \cellcolor[HTML]{92C47E}15.7 &
  \cellcolor[HTML]{A9C879}21.7 &
  \cellcolor[HTML]{D0CE70}31.5 &
  \cellcolor[HTML]{B9CA75}25.7 &
  \cellcolor[HTML]{8AC380}13.7 &
  \cellcolor[HTML]{9BC67C}18.2 &
  \cellcolor[HTML]{E67C73}68.7 &
  \cellcolor[HTML]{E67C73}69.5 &
  \cellcolor[HTML]{E67C73}68.1 &
  \cellcolor[HTML]{E67C73}70.3 &
  \cellcolor[HTML]{E67C73}60.0 &
  \cellcolor[HTML]{E67C73}55.2 &
  \cellcolor[HTML]{E67C73}68.7 &
  \cellcolor[HTML]{E67C73}76.2 \\ \cline{2-2}
 &
  F1 &
  \cellcolor[HTML]{D6CF6F}32.9 &
  \cellcolor[HTML]{CECE71}31.0 &
  \cellcolor[HTML]{C8CD72}29.5 &
  \cellcolor[HTML]{E6D26C}37.0 &
  \cellcolor[HTML]{C6CC73}29.0 &
  \cellcolor[HTML]{D5CF6F}32.7 &
  \cellcolor[HTML]{8CC37F}14.2 &
  \cellcolor[HTML]{DCD06E}34.4 &
  \cellcolor[HTML]{DDD06E}34.8 &
  \cellcolor[HTML]{DCD06E}34.3 &
  \cellcolor[HTML]{E67C73}50.7 &
  \cellcolor[HTML]{E88272}49.6 &
  \cellcolor[HTML]{C6CC73}28.8 &
  \cellcolor[HTML]{A4C77A}20.3 &
  \cellcolor[HTML]{DAD06E}33.8 &
  \cellcolor[HTML]{E0D16D}35.6 \\ \cline{1-2}
\multirow{-3}{*}{A07} &
  F20 &
  \cellcolor[HTML]{98C57D}17.2 &
  \cellcolor[HTML]{98C57D}17.2 &
  \cellcolor[HTML]{9EC67B}18.8 &
  \cellcolor[HTML]{C3CC73}28.2 &
  \cellcolor[HTML]{E2D16D}36.1 &
  \cellcolor[HTML]{CECE71}30.9 &
  \cellcolor[HTML]{A2C77A}19.9 &
  \cellcolor[HTML]{9AC57C}17.7 &
  \cellcolor[HTML]{E67C73}77.7 &
  \cellcolor[HTML]{E67C73}79.0 &
  \cellcolor[HTML]{E67C73}69.2 &
  \cellcolor[HTML]{E67C73}70.9 &
  \cellcolor[HTML]{E67C73}60.8 &
  \cellcolor[HTML]{E67C73}89.0 &
  \cellcolor[HTML]{E67C73}67.5 &
  \cellcolor[HTML]{E67C73}72.7 \\ \cline{2-2}
 &
  F66 &
  \cellcolor[HTML]{86C280}12.8 &
  \cellcolor[HTML]{86C280}12.7 &
  \cellcolor[HTML]{92C47E}15.7 &
  \cellcolor[HTML]{A9C879}21.7 &
  \cellcolor[HTML]{D0CE70}31.5 &
  \cellcolor[HTML]{B9CA75}25.7 &
  \cellcolor[HTML]{8AC380}13.7 &
  \cellcolor[HTML]{9BC67C}18.2 &
  \cellcolor[HTML]{E67C73}68.7 &
  \cellcolor[HTML]{E67C73}69.5 &
  \cellcolor[HTML]{E67C73}68.1 &
  \cellcolor[HTML]{E67C73}70.3 &
  \cellcolor[HTML]{E67C73}60.0 &
  \cellcolor[HTML]{E67C73}55.2 &
  \cellcolor[HTML]{E67C73}68.7 &
  \cellcolor[HTML]{E67C73}76.2 \\ \cline{1-2}
\multirow{-3}{*}{A08} &
  F86 &
  \cellcolor[HTML]{C7CD72}26.3 &
  \cellcolor[HTML]{B5CA76}22.3 &
  \cellcolor[HTML]{BDCB75}23.9 &
  \cellcolor[HTML]{D0CE71}28.2 &
  \cellcolor[HTML]{BACA75}23.3 &
  \cellcolor[HTML]{BECB74}24.2 &
  \cellcolor[HTML]{AEC978}20.6 &
  \cellcolor[HTML]{E77E72}49.8 &
  \cellcolor[HTML]{77C084} 8.0 &
  \cellcolor[HTML]{74BF84} 7.3 &
  \cellcolor[HTML]{E67C73}61.5 &
  \cellcolor[HTML]{FBD567}38.0 &
  \cellcolor[HTML]{E98571}49.0 &
  \cellcolor[HTML]{E67C73}65.0 &
  \cellcolor[HTML]{F1A16D}45.5 &
  \cellcolor[HTML]{F5B26B}43.3 \\ \cline{2-2}
 &
  F85 &
  \cellcolor[HTML]{FCD567}38.3 &
  \cellcolor[HTML]{FDCD67}40.0 &
  \cellcolor[HTML]{F3D469}36.2 &
  \cellcolor[HTML]{FCD567}38.3 &
  \cellcolor[HTML]{F3D469}36.2 &
  \cellcolor[HTML]{F9D568}37.5 &
  \cellcolor[HTML]{D6CF6F}29.5 &
  \cellcolor[HTML]{E67C73}53.9 &
  \cellcolor[HTML]{86C280}11.5 &
  \cellcolor[HTML]{82C281}10.7 &
  \cellcolor[HTML]{E67C73}63.2 &
  \cellcolor[HTML]{ECD36A}34.7 &
  \cellcolor[HTML]{E67C73}68.0 &
  \cellcolor[HTML]{E67C73}97.0 &
  \cellcolor[HTML]{EE976F}46.8 &
  \cellcolor[HTML]{F2A56D}45.0 \\ \cline{1-2}
\multirow{-3}{*}{A09} &
  F86 &
  \cellcolor[HTML]{C7CD72}26.3 &
  \cellcolor[HTML]{B5CA76}22.3 &
  \cellcolor[HTML]{BDCB75}23.9 &
  \cellcolor[HTML]{D0CE71}28.2 &
  \cellcolor[HTML]{BACA75}23.3 &
  \cellcolor[HTML]{BECB74}24.2 &
  \cellcolor[HTML]{AEC978}20.6 &
  \cellcolor[HTML]{E77E72}49.8 &
  \cellcolor[HTML]{77C084} 8.0 &
  \cellcolor[HTML]{74BF84} 7.3 &
  \cellcolor[HTML]{E67C73}61.5 &
  \cellcolor[HTML]{FBD567}38.0 &
  \cellcolor[HTML]{E98571}49.0 &
  \cellcolor[HTML]{E67C73}65.0 &
  \cellcolor[HTML]{F1A16D}45.5 &
  \cellcolor[HTML]{F2A56D}45.0 \\ \cline{2-2}
 &
  F85 &
  \cellcolor[HTML]{FCD567}38.3 &
  \cellcolor[HTML]{FDCD67}40.0 &
  \cellcolor[HTML]{F3D469}36.2 &
  \cellcolor[HTML]{FCD567}38.3 &
  \cellcolor[HTML]{F3D469}36.2 &
  \cellcolor[HTML]{F9D568}37.5 &
  \cellcolor[HTML]{D6CF6F}29.5 &
  \cellcolor[HTML]{E67C73}53.9 &
  \cellcolor[HTML]{86C280}11.5 &
  \cellcolor[HTML]{82C281}10.7 &
  \cellcolor[HTML]{E67C73}63.2 &
  \cellcolor[HTML]{ECD36A}34.7 &
  \cellcolor[HTML]{E67C73}68.0 &
  \cellcolor[HTML]{E67C73}97.0 &
  \cellcolor[HTML]{EE976F}46.8 &
  \cellcolor[HTML]{F5B26B}43.3 \\ \cline{1-2}
\multirow{-3}{*}{A10} &
  F58 &
  \cellcolor[HTML]{EA8871}48.6 &
  \cellcolor[HTML]{E67C73}50.3 &
  \cellcolor[HTML]{E88372}49.3 &
  \cellcolor[HTML]{E88272}49.3 &
  \cellcolor[HTML]{F3AA6C}44.3 &
  \cellcolor[HTML]{F7B76A}42.7 &
  \cellcolor[HTML]{FAC169}41.4 &
  \cellcolor[HTML]{EB8D70}48.0 &
  \cellcolor[HTML]{F4D469}36.5 &
  \cellcolor[HTML]{FCD567}38.2 &
  \cellcolor[HTML]{97C57D}15.3 &
  \cellcolor[HTML]{E67C73}51.3 &
  \cellcolor[HTML]{E67C73}50.8 &
  \cellcolor[HTML]{E67C73}54.8 &
  \cellcolor[HTML]{FFD566}38.9 &
  \cellcolor[HTML]{E77F72}49.7 \\ \cline{2-2}
 &
  F60 &
  \cellcolor[HTML]{EF9B6E}46.2 &
  \cellcolor[HTML]{EB8D70}48.0 &
  \cellcolor[HTML]{EF9D6E}46.0 &
  \cellcolor[HTML]{E98671}48.8 &
  \cellcolor[HTML]{F5B16B}43.5 &
  \cellcolor[HTML]{FBC768}40.7 &
  \cellcolor[HTML]{FCCB67}40.3 &
  \cellcolor[HTML]{EF9B6E}46.3 &
  \cellcolor[HTML]{F0D36A}35.6 &
  \cellcolor[HTML]{FAD567}37.8 &
  \cellcolor[HTML]{97C57D}15.3 &
  \cellcolor[HTML]{E67C73}52.6 &
  \cellcolor[HTML]{E98571}49.0 &
  \cellcolor[HTML]{E67C73}58.7 &
  \cellcolor[HTML]{F8D468}37.3 &
  \cellcolor[HTML]{FBC768}40.7 \\ \cline{1-2}
\multirow{-3}{*}{A11} &
  F82 &
  \cellcolor[HTML]{EAD26B}34.1 &
  \cellcolor[HTML]{E7D26B}33.6 &
  \cellcolor[HTML]{F3D469}36.2 &
  \cellcolor[HTML]{D5CF6F}29.3 &
  \cellcolor[HTML]{D6CF6F}29.6 &
  \cellcolor[HTML]{D2CE70}28.7 &
  \cellcolor[HTML]{EC9070}47.6 &
  \cellcolor[HTML]{E67C73}60.2 &
  \cellcolor[HTML]{F5D469}36.6 &
  \cellcolor[HTML]{EBD26B}34.3 &
  \cellcolor[HTML]{E67C73}64.5 &
  \cellcolor[HTML]{A3C77A}18.2 &
  \cellcolor[HTML]{E67C73}66.1 &
  \cellcolor[HTML]{E67C73}94.0 &
  \cellcolor[HTML]{F6B36B}43.2 &
  \cellcolor[HTML]{E67C73}51.3 \\ \cline{2-2}
 &
  F19 &
  \cellcolor[HTML]{E67C73}86.2 &
  \cellcolor[HTML]{E67C73}87.0 &
  \cellcolor[HTML]{E67C73}84.7 &
  \cellcolor[HTML]{E67C73}83.3 &
  \cellcolor[HTML]{E67C73}80.6 &
  \cellcolor[HTML]{E67C73}82.9 &
  \cellcolor[HTML]{E67C73}82.2 &
  \cellcolor[HTML]{E67C73}77.0 &
  \cellcolor[HTML]{FBD567}38.1 &
  \cellcolor[HTML]{F8D468}37.3 &
  \cellcolor[HTML]{F3AA6C}44.3 &
  \cellcolor[HTML]{B1C977}21.3 &
  \cellcolor[HTML]{F1A26D}45.3 &
  \cellcolor[HTML]{F1A16D}45.5 &
  \cellcolor[HTML]{EDD36A}34.8 &
  \cellcolor[HTML]{E6D26C}33.3 \\ \cline{1-2}
\multirow{-3}{*}{A12} &
  F67 &
  \cellcolor[HTML]{E1D16D}32.0 &
  \cellcolor[HTML]{DAD06E}30.6 &
  \cellcolor[HTML]{EAD26B}34.1 &
  \cellcolor[HTML]{C2CC74}25.0 &
  \cellcolor[HTML]{C0CB74}24.6 &
  \cellcolor[HTML]{C3CC73}25.3 &
  \cellcolor[HTML]{C1CC74}24.8 &
  \cellcolor[HTML]{FFD566}39.0 &
  \cellcolor[HTML]{FBC768}40.7 &
  \cellcolor[HTML]{FECF67}39.7 &
  \cellcolor[HTML]{E6D26C}33.2 &
  \cellcolor[HTML]{E88172}49.5 &
  \cellcolor[HTML]{B5CA76}22.1 &
  \cellcolor[HTML]{95C57D}15.0 &
  \cellcolor[HTML]{FBD567}38.1 &
  \cellcolor[HTML]{F1D369}35.8 \\ \cline{2-2}
 &
  F39 &
  \cellcolor[HTML]{F3D469}36.1 &
  \cellcolor[HTML]{E6D26C}33.3 &
  \cellcolor[HTML]{F8D468}37.3 &
  \cellcolor[HTML]{F7B96A}42.5 &
  \cellcolor[HTML]{E8D26B}33.7 &
  \cellcolor[HTML]{E7D26C}33.4 &
  \cellcolor[HTML]{C5CC73}25.8 &
  \cellcolor[HTML]{FBD567}37.9 &
  \cellcolor[HTML]{FDD567}38.4 &
  \cellcolor[HTML]{FCC967}40.5 &
  \cellcolor[HTML]{D2CE70}28.7 &
  \cellcolor[HTML]{E88272}49.3 &
  \cellcolor[HTML]{B8CA76}22.8 &
  \cellcolor[HTML]{ACC878}20.1 &
  \cellcolor[HTML]{F6D468}36.8 &
  \cellcolor[HTML]{E5D16C}33.0 \\ \cline{1-2}
\multirow{-3}{*}{A13} &
  F45 &
  \cellcolor[HTML]{E67C73}52.8 &
  \cellcolor[HTML]{E67C73}53.3 &
  \cellcolor[HTML]{E67C73}56.9 &
  \cellcolor[HTML]{EF9C6E}46.7 &
  \cellcolor[HTML]{E67C73}50.4 &
  \cellcolor[HTML]{EC8F70}48.1 &
  \cellcolor[HTML]{E67C73}71.2 &
  \cellcolor[HTML]{E67C73}55.6 &
  \cellcolor[HTML]{E67C73}80.9 &
  \cellcolor[HTML]{E67C73}80.8 &
  \cellcolor[HTML]{FCD567}39.9 &
  \cellcolor[HTML]{FCCA67}41.8 &
  \cellcolor[HTML]{DBD06E}32.1 &
  \cellcolor[HTML]{57BB8A} 0.8 &
  \cellcolor[HTML]{F2A66C}45.6 &
  \cellcolor[HTML]{E67C73}55.0 \\ \cline{2-2}
 &
  F51 &
  \cellcolor[HTML]{E67C73}51.0 &
  \cellcolor[HTML]{E67C73}51.0 &
  \cellcolor[HTML]{E67C73}54.3 &
  \cellcolor[HTML]{F4AD6B}44.8 &
  \cellcolor[HTML]{F3AA6C}45.2 &
  \cellcolor[HTML]{FBC668}42.3 &
  \cellcolor[HTML]{E67C73}65.0 &
  \cellcolor[HTML]{E67C73}53.2 &
  \cellcolor[HTML]{E67C73}73.8 &
  \cellcolor[HTML]{E67C73}73.8 &
  \cellcolor[HTML]{E5D16C}34.3 &
  \cellcolor[HTML]{F9BF69}43.0 &
  \cellcolor[HTML]{CFCE71}29.3 &
  \cellcolor[HTML]{57BB8A} 0.8 &
  \cellcolor[HTML]{FBC868}42.0 &
  \cellcolor[HTML]{E67C73}56.0 \\ \cline{1-2}
\multirow{-3}{*}{A14} &
  F78 &
  \cellcolor[HTML]{E67C73}56.2 &
  \cellcolor[HTML]{E67C73}59.7 &
  \cellcolor[HTML]{E67C73}57.7 &
  \cellcolor[HTML]{E67C73}55.7 &
  \cellcolor[HTML]{E67C73}52.2 &
  \cellcolor[HTML]{E67C73}53.2 &
  \cellcolor[HTML]{E67C73}57.8 &
  \cellcolor[HTML]{EA8A71}49.0 &
  \cellcolor[HTML]{FFD666}43.2 &
  \cellcolor[HTML]{FAC368}44.7 &
  \cellcolor[HTML]{F7D468}41.2 &
  \cellcolor[HTML]{E67C73}51.8 &
  \cellcolor[HTML]{E67C73}62.7 &
  \cellcolor[HTML]{E67C73}58.5 &
  \cellcolor[HTML]{A5C77A}20.7 &
  \cellcolor[HTML]{F8BA6A}45.3 \\ \cline{2-2}
 &
  F77 &
  \cellcolor[HTML]{E88272}49.6 &
  \cellcolor[HTML]{E78072}49.8 &
  \cellcolor[HTML]{E67C73}50.3 &
  \cellcolor[HTML]{EB8C70}48.8 &
  \cellcolor[HTML]{EE976F}48.0 &
  \cellcolor[HTML]{EF9C6E}47.6 &
  \cellcolor[HTML]{E67C73}55.3 &
  \cellcolor[HTML]{F8BC69}45.2 &
  \cellcolor[HTML]{ED956F}48.2 &
  \cellcolor[HTML]{EA8B70}48.9 &
  \cellcolor[HTML]{ADC878}22.6 &
  \cellcolor[HTML]{F5B16B}46.0 &
  \cellcolor[HTML]{E67C73}58.3 &
  \cellcolor[HTML]{D0CE71}31.3 &
  \cellcolor[HTML]{B9CA75}25.7 &
  \cellcolor[HTML]{FDCE67}43.8 \\ \cline{1-2}
\multirow{-3}{*}{A15} &
  F66 &
  \cellcolor[HTML]{E67C73}87.2 &
  \cellcolor[HTML]{E67C73}87.3 &
  \cellcolor[HTML]{E67C73}84.3 &
  \cellcolor[HTML]{E67C73}78.3 &
  \cellcolor[HTML]{E67C73}68.5 &
  \cellcolor[HTML]{E67C73}74.3 &
  \cellcolor[HTML]{E67C73}86.3 &
  \cellcolor[HTML]{E67C73}81.8 &
  \cellcolor[HTML]{DED06E}31.3 &
  \cellcolor[HTML]{DAD06E}30.5 &
  \cellcolor[HTML]{E0D16D}31.9 &
  \cellcolor[HTML]{D6CF6F}29.7 &
  \cellcolor[HTML]{FDCD67}40.0 &
  \cellcolor[HTML]{F2A66D}44.8 &
  \cellcolor[HTML]{DED06E}31.3 &
  \cellcolor[HTML]{BCCB75}23.8 \\ \cline{2-2}
 \multirow{-2}{*}{A16} &
  F18 &
  \cellcolor[HTML]{E67C73}81.0 &
  \cellcolor[HTML]{E67C73}82.2 &
  \cellcolor[HTML]{E67C73}79.2 &
  \cellcolor[HTML]{E67C73}73.7 &
  \cellcolor[HTML]{E67C73}69.2 &
  \cellcolor[HTML]{E67C73}73.8 &
  \cellcolor[HTML]{E67C73}77.1 &
  \cellcolor[HTML]{E67C73}76.5 &
  \cellcolor[HTML]{DDD06E}31.2 &
  \cellcolor[HTML]{D5CF6F}29.5 &
  \cellcolor[HTML]{F8BA6A}42.3 &
  \cellcolor[HTML]{BFCB74}24.3 &
  \cellcolor[HTML]{EF9B6E}46.3 &
  \cellcolor[HTML]{FAD567}37.8 &
  \cellcolor[HTML]{EBD26B}34.4 &
  \cellcolor[HTML]{CACD72}26.8 \\ \cline{1-2}
\end{tabular}%
}
\label{tab:expB}
\end{table*}

%% file: tex/experiments/expC.tex
\subsection{Experiment C: Comparision with Neural Front-end}

%% file: tex/results/expC_single.tex
\begin{figure}
    \centering
    \includegraphics[width=\linewidth,trim={0 0 0 0},clip]{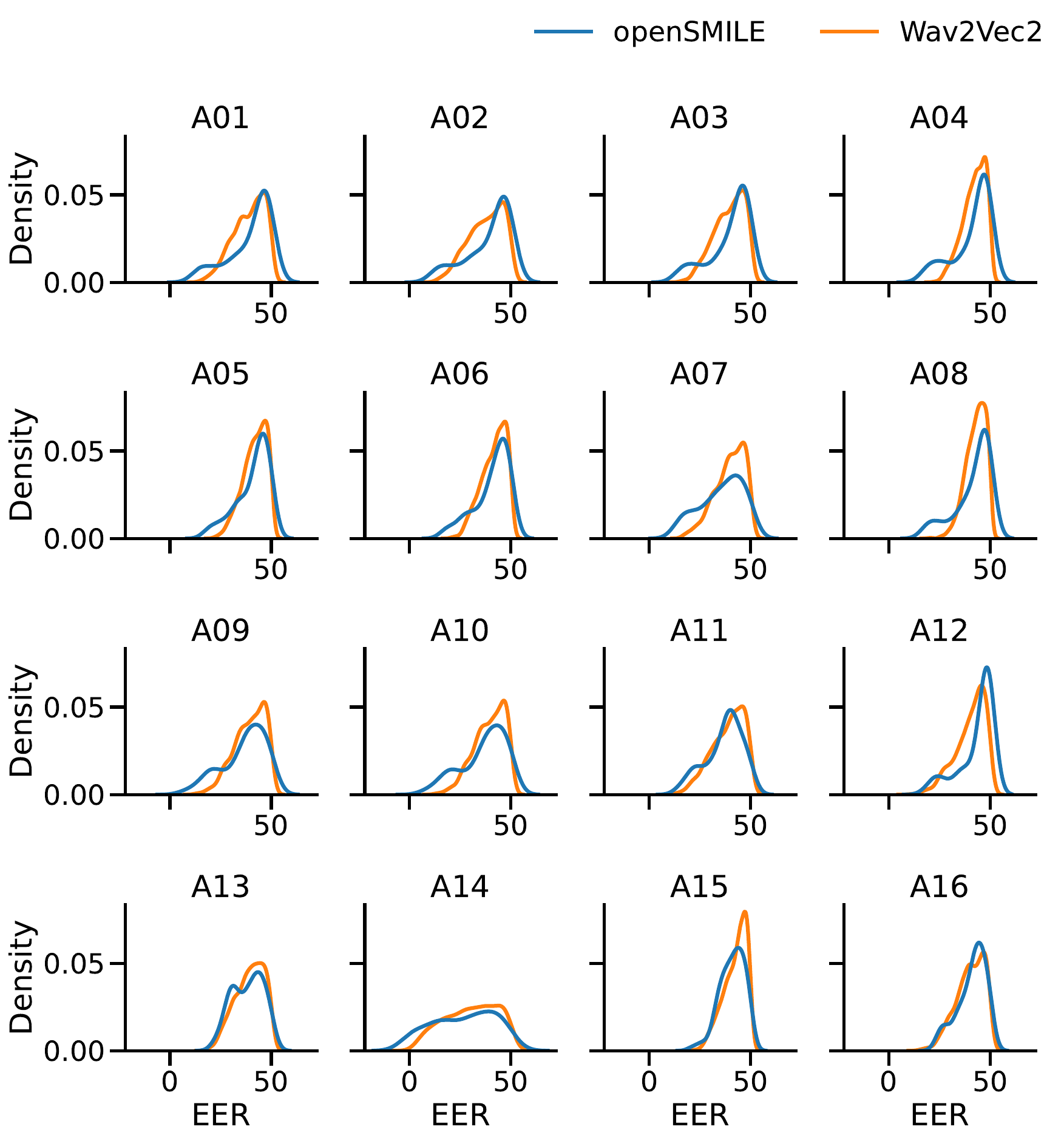}
    \caption{%
    Distribution plot comparing the performance of scalar-valued features extracted using openSMILE (blue) and Wav2Vec2 (orange) across  attacks $A01$ to $A16$. 
    The $x$-axis represents the EER obtained by each feature, and the $y$-axis denotes frequency. 
    The plot reveals that openSMILE features sometimes exhibit higher predictive accuracy, as evidenced by a greater concentration of distributional mass near an $EER=0$.
    }
    \label{fig:openSMILE-vs-Wav2Vec2-single-feature}
\end{figure}

We compare the `eGeMAPSv2' openSMILE front-end to the Wav2Vec2\footnote{\url{huggingface.co/facebook/wav2vec2-base}} self-supervised neural front-end \cite{baevski20Wav2Vec2},
used widely in related work \cite{wang22odyssey,donas22icassp,tak2022odyssey,yang2024icassp,guo2024icassp,pascu24interspeech}.
For each of the 768 scalar outputs of Wav2Vec2, we assess how discriminative it is for each of the sixteen attacks in ASV5.
This is done by using the scalar value as a score itself and directly computing the EER on them, similar to \Cref{fig:dist}.
Thus, for each attack, we obtain 768 predictors with corresponding EERs.
We perform the same process for openSMILE and compare via a (normalized) distribution plot, c.f. \Cref{fig:openSMILE-vs-Wav2Vec2-single-feature}.
We observe that for both openSMILE and Wav2Vec2, many of the features are not predictive: the performance has its mode around 40--50\% EER.
However, for both feature sets there is a small subset of more predictive features (around 20--30\% EER).
Note that this subset is much more distinct for openSMILE, which tends to exhibit a bimodal distribution for some attacks (e.g., $A04$, $A08$, $A09$, $A13$).
Note also the attack $A14$, unlike other attacks, seems to be easily detectable by many of its features, as indicated by the high distributional mass at EER = 0.

%% file: tex/results/expC_multi.tex

\begin{table}[ht]
    \centering
    \caption{
    Comparison of aggregated ID and OOD performance between openSMILE and Wav2Vec2 features.
    }
    \begin{tabular}{c||c|c||c|c}
    & \multicolumn{2}{c||}{In-domain test EER} & \multicolumn{2}{c}{Out-of-domain test EER} \\
    Attack & openSMILE & W2V2 & openSMILE & W2V2 \\ \hline
    A01 &           1.6 &          0.0 &           41.9 &           5.5 \\
    A02 &           0.6 &          0.0 &           40.2 &           4.8 \\
    A03 &           1.7 &          0.0 &           38.9 &           4.7 \\
    A04 &           1.4 &          0.3 &           33.9 &          14.3 \\
    A05 &           1.0 &          0.3 &           26.7 &          15.6 \\
    A06 &           1.3 &          0.2 &           27.5 &          13.5 \\
    A07 &           0.3 &          0.0 &           37.2 &          13.1 \\
    A08 &           5.0 &          0.0 &           41.6 &           9.3 \\
    \hline
    A09 &           2.6 &          0.0 &           34.7 &          13.1 \\
    A10 &           2.1 &          0.2 &           35.2 &          11.1 \\
    A11 &           0.7 &          0.0 &           30.7 &          14.8 \\
    A12 &           8.5 &          0.2 &           50.1 &          25.3 \\
    A13 &           2.5 &          0.5 &           33.7 &          16.0 \\
    A14 &           0.0 &          0.0 &           38.5 &          18.8 \\
    A15 &           6.5 &          0.5 &           41.6 &           8.1 \\
    A16 &           3.3 &          0.0 &           33.8 &           9.1 
    \end{tabular}
    \label{tab:openSMILE-vs-Wav2Vec2-all-features}
\end{table}

We train a logistic regression classifier on all features for both openSMILE and Wav2Vec2.
Similar to the experiments in \Cref{subsec:exp-b}, we consider all combinations of train--test attacks, 
that is, we train on one attack and evaluate on all others.
However, for brevity, we report only aggregated results for the two main cases:
in domain (when train and test attacks match), and
out of domain (when train and test attacks do not match).
The results are shown in \Cref{tab:openSMILE-vs-Wav2Vec2-all-features}.
Here, we observe the reverse trend to the previous experiment:
Wav2Vec2 gives better performance than openSMILE for both in-domain and out-of-domain configuration.
Thus, while individual features of Wav2Vec2 are not as predictive,
they do perform much better as a whole.
We also see that using all features for openSMILE significantly improves performance over the single features in both \Cref{tab:expA} and \Cref{tab:expB}.


%% file: tex/related_work.tex
Kumar \etal \cite{kumar2022fake} use openSMILE's `eGeMAPSv2' as a front-end, combined with traditional machine learning models such as decision trees or linear regression, and evaluate on three datasets: CMU Arctic \cite{kominek2004arctic}, LibriTTS \cite{zen2019libritts}, and the LJ Speech \cite{ljspeech17} dataset. 
The corresponding spoofed audio files are created by themselves.
Their results support our finding that in an ID setting, openSMILE front-ends work very reliably, while OOD performance degenerates to random guessing.
Similar results are obtained by Sadashiv \etal~\cite{tn2024fake}, who evaluate on 
ASVspoof 2019 \cite{asvspoof19} in addition to  CMU Arctic, LJ Speech, and LibriTTS.
Other interpretable features such as energy, distortion, loudness have been used by M\"uller \etal \cite{muller22attacker} in the context of source tracing.
Furthermore, frequency-based features such as fundamental frequency (F0) \cite{xue22f0} and frequency sub-band \cite{tak20cqcc,yang20tifs,salvi2023eusipco,salvi2024listening} have been shown to be useful.
Finally, a signal's noise (high-frequency information) 
 can also be used for voice anti-spoofing\cite{salvi2024listening}.
